\documentclass[aps,pre,twocolumn,superscriptaddress,showpacs,showkeys]{revtex4}
\usepackage{epsfig}
\usepackage{graphics,amsmath,amssymb}
\usepackage[english]{babel}

\begin{document}
\title{Revealing the phase transition behaviors of k-core percolation in random networks}
\author{Yong Zhu}
\author{Xiaosong Chen}
\affiliation{State Key Laboratory of Theoretical Physics, Institute of Theoretical Physics, Chinese Academy of Sciences, P.O. Box 2735, Beijing 100190, China}

\date{\today}

\begin{abstract}
The $k$-core percolation is a fundamental structural transition in complex networks. Through the analysis of the size jump behaviors of $k$-core in the evolution process of networks, we confirm that $k$-core percolation is continuous phase transition when $k=1,2$ while it is a hybrid first-order-second-order phase transition when $k\ge 3$. $2$-core percolation belongs to different universality class from that of $1$-core (giant component) percolation. The discontinuity of $k$-core percolation with $k\ge 3$ can be concluded from largest size jump of $k$-core which will not disappear in the thermodynamic limit while its continuous characteristic is reflected by second largest size jump which converges to zero in power law as $N\to \infty$. Furthermore, along with the previously known exponent $\beta=0.5$, we obtain a set of exponents which are independent of $k$ when $k\ge 3$ and also different from those critical exponents of $1$-core and $2$-core percolation.

\end{abstract}

\maketitle

\section{Introduction}

The $k$-core~\cite{Bollobas84} of a network is its largest subgraph each of whose vertices has at least k nearest neighbors within this subgraph. $k$-core (bootstrap) percolation which implies the emergence of the giant $k$-core when edges are added randomly into the networks was first introduced on Bethe networks by J.Chalupa {\it et al}.\cite{Chalupa79}

When $k=1$, $1$-core percolation leads to well studied percolation of the giant component which is a continuous phase transition occurs at a critical connecting probability $p_c$. When $p\ge p_c$, the relative size of giant component satisfies $s\propto (p-p_c)^{\beta}$ where $\beta=1$ according to the mean-field theory. The $2$-core is similar but note identical to the bi-connected component of a network since a bi-connected component is surely a $2$-core while the reverse is not. $2$-core percolation is also continuous phase transition with the same critical point as percolation of giant component but the critical exponent $\beta=2$.

In contrast to $k$-core percolation with $k=1,2$, when $k\ge 3$, the $k$-core appears suddenly at a transition point $p_c$. When $p\ge p_c$, the relative size of $k$-core increases as
\begin{equation}
s(p)-s(p_c)\propto (p-p_c)^{\beta}
\label{eq:hybrid}
\end{equation}
where $\beta=1/2$ and $s(p_c)$ is a nonzero constant which increases with $k$. It indicates that the order parameter has a jump at $p = p_c$ from 0 to $s(p_c)$ as at an ordinary first-order phase transition as well as a singular behaviour as at a continuous phase transition~\cite{Stauffer94}. Thus $k$-core percolation with $k\ge 3$ on Bethe networks is a hybrid phase transition.

This intriguing hybrid characteristic is also discovered for $k$-core percolation in Erd\"os-R\'enyi (ER) networks\cite{Pittel96} and configuration model with arbitrary degree distribution~\cite{Dorogovtsev06a,Dorogovtsev06b,Goltsev06}, heterogeneous $k$-core percolation~\cite{Baxter11}, core percolation in directed networks where the in- and out-degree distributions differ~\cite{Liuyy12} and a two parameter-dependent $(K,K^{'})-$protected core percolation~\cite{Zhaojh13}.

It is worth mentioning that the hybrid nature of $k$-core percolation expressed as Eq.\eqref{eq:hybrid} has not been revealed through simulation method yet. In this paper, we will use Monte Carlo simulation to investigate the $k$-core in ER networks and illustrate its phase transition properties with the analysis of the size jump behaviors of $k$-core in the evolution process of networks. The rest of this paper is organized as follows. In Sec.\ref{sec:model_method}, we introduce $k$-core percolation in ER random networks and method to be used to analyze the simulation data. Our results of continuous $k$-core percolation ($k=1,2$) and hybrid $k$-core percolation ($k\ge 3$) are presented in Sec.\ref{sec:result_k12} and \ref{sec:result_k345} respectively. Finally a summarization is given in Sec.\ref{sec:summary}.

\section{Model and Method}
\label{sec:model_method}
$k$-core percolation in ER random networks was first studied by Pittel et al. through probabilistic analysis with transition point $p_c$ and $s(p_c)$ obtained~\cite{Pittel96}. The $k$-core of a specific random network can be obtained in the following way. Remove all vertices of degree less than $k$ from the network. Some of the rest of the vertices may remain with less than k edges. Then remove these vertices, and so on, until no further removal is possible. It has been pointed out by Pittel {\it et al.} that, for ER random networks, the remaining vertices after pruning process, if there are, belong to a single $k$-core~\cite{Pittel96}. Accordingly, the $k$-core percolation in the process of adding edges randomly to networks can be simply solved by deleting edges in the exact reverse order. In each step, if the edge to be deleted have already been removed in previous pruning process, then move to the next step. Otherwise, pruning process is carried out to find the new $k$-core.

The relative size $s$ of the largest cluster after sample averaging is usually taken as the order parameter of percolation. Controlling parameter $r$ is defined as the edge number reduced by system size $N$. For continuous percolation, $s$ along with the sizes of other ranked clusters changes continuously with $r$ and shows power-law decay with system size $N$ at the critical point $r_c$. Besides the sizes of ranked clusters themselves, percolation phase transitions can be characterized by the size jump behaviors of largest cluster in each realization of networks\cite{Lee11,Nagler11,Qian12,Fan14,Zhu15a,Zhu15b}. Here, we define sample-dependent pseudo-critical point as the $r_c^{i}$ where order parameter exhibits a sudden biggest jump in $i$-th run and the corresponding size gap is denoted as $\Delta^{(i)}$. For a network of each size $N$, $512,000$ runs are made. The averages of $r_c^{i}$ and $\Delta^{(i)}$ are denoted respectively as $\bar{r}_c (N)$ and $\bar{\Delta}(N)$. Their sample-to-sample fluctuations are calculated as
\begin{eqnarray}
\chi_r &\equiv \sqrt{<(r_c^{(i)})^2>-\bar{r}_c (N)}, \\
\chi_\Delta &\equiv \sqrt{<(\Delta^{(i)})^2>-\bar{\Delta}(N)}.
\end{eqnarray}

The following finite size scaling hypotheses are made and confirmed by simulation data for continuous percolation~\cite{Fan14,Zhu15a,Zhu15b}
\begin{align}
\bar{r}_c (N)&= r_c (\infty)+a_r N^{-(1/\nu)_1}, \label{eq_r_c} \\
\bar{\Delta} (N)&= a_\Delta N^{-(\beta/\nu)_1},  \label{eq_delta}\\
\chi_r &= b_r N^{-(1/\nu)_2},   \label{eq_chi_r}\\
\chi_\Delta &= b_{\Delta} N^{-(\beta/\nu)_2}. \label{eq_chi_delta}
\end{align}
where $1/\nu_1$ and $1/\nu_2$ are not always equal while $(\beta/\nu)_1$ and $(\beta/\nu)_2$ are. The asymptotic behavior of $\bar{\Delta} (N)$ was used by Nagler {\it et al.} to evaluate the extent of discontinuity of percolation phase transition~\cite{Nagler11}. However, the so-called weakly discontinuous percolation phase transitions are actually continuous.

It's anticipated for continuous percolation that not only the largest but also the second, third... largest size jump of largest cluster and the corresponding sample-dependent pseudo-critical points are related to percolation phase transition and satisfy the finite size scaling hypotheses above.

As for discontinuous percolation, the averaged size jump $\bar{\Delta} (N)$ is intuitively expected to converge to a nonvanishing constant as the system size $N\to \infty$ while the asymptotic behaviors of the three other quantities are unknown.

\begin{figure}[h!]
\resizebox{0.48\textwidth}{!}{\includegraphics{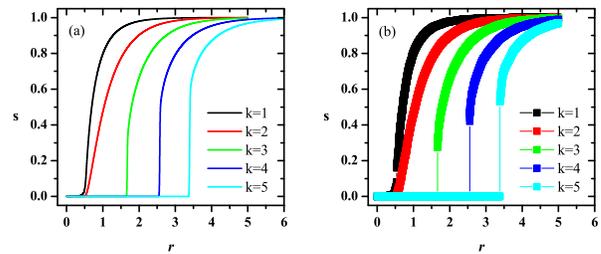}}
\caption{(a)Plots of order parameter $s$ of $k-$core percolation versus $r$. (b)Evolutions of the $k$-core sizes versus $r$ in the same single realization of networks. The system size is $N=10000$.}
\label{fig:1}
\end{figure}

For $k$-core percolation, the relative size of $k$-core is naturally taken as order parameter. In Fig.\ref{fig:1}(a) the order parameter of $k-$core percolation in networks with $N=10000$ is plotted against $r$. Apparently, $s$ increases continuously in the cases of $k=1,2$ but increases sharply in the cases of $k=3,4,5$.

In Fig.\ref{fig:1}(b), evolutions of the sizes of $k$-core with different $k$ in the same single realization are displayed. In particular, when $k=3,4,5$, $k$-core remains absent and becomes macroscopic suddenly, indicating the discontinuity of percolation phase transitions. The largest jump size of $1$-core is much smaller and that of $2-$core is even invisible. Actually, there are many small size jumps along with the largest one.

It's of great interest to investigate the asymptotic properties of size jump behaviors of $k$-core. The properties of largest and second largest size jumps are involved in this paper. Their corresponding quantities are marked with superscripts 1 and 2 respectively for the sake of distinction.

\section{The case of continuous $k$-core percolation ($k=1,2$)}
\label{sec:result_k12}
\begin{figure}[h!]
\resizebox{0.48\textwidth}{!}{\includegraphics{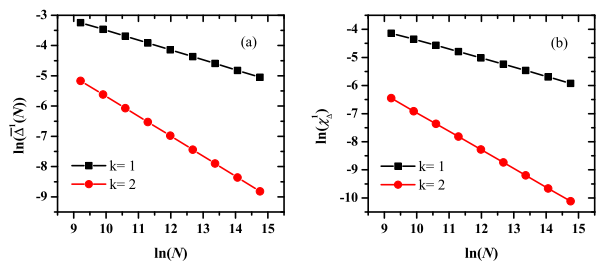}}
\caption{Asymptotic behaviors of the largest size gaps of $k$-core with $k=1,2$. (a) $\ln(\bar{\Delta}^1 (N))$ as a function of $\ln(N)$; (b) $\ln(\chi_{\Delta}^1)$ as a function of $\ln(N)$. Both $(\beta/\nu)_1^1$ and $(\beta/\nu)_2^1$ equal approximately $1/3$ for $k=1$ and $2/3$ for $k=2$. }
\label{fig:2}
\end{figure}

It's known that in ER networks percolation of $1$-core and $2$-core undergoes continuous transitions at the same critical point $r_c=0.5$. In Fig.\ref{fig:2}, we show the asymptotic behaviors of the averages and fluctuations of the largest size gaps of $k$-core with $k=1,2$. As expected, power-law behaviors are found for both $\bar{\Delta}^1 (N)$ and $\chi_{\Delta}^1$. It can be seen clearly that both $(\beta/\nu)_1^1$ and $(\beta/\nu)_2^1$ for $k=1$ are different from those for $k=2$. Nevertheless, $(\beta/\nu)_1^1$ remains approximately equal to $(\beta/\nu)_2^1$ within the error range. Specifically, we get $1/3$ for $k=1$ and $2/3$ for $k=2$.

\begin{figure}[h!]
\resizebox{0.48\textwidth}{!}{\includegraphics{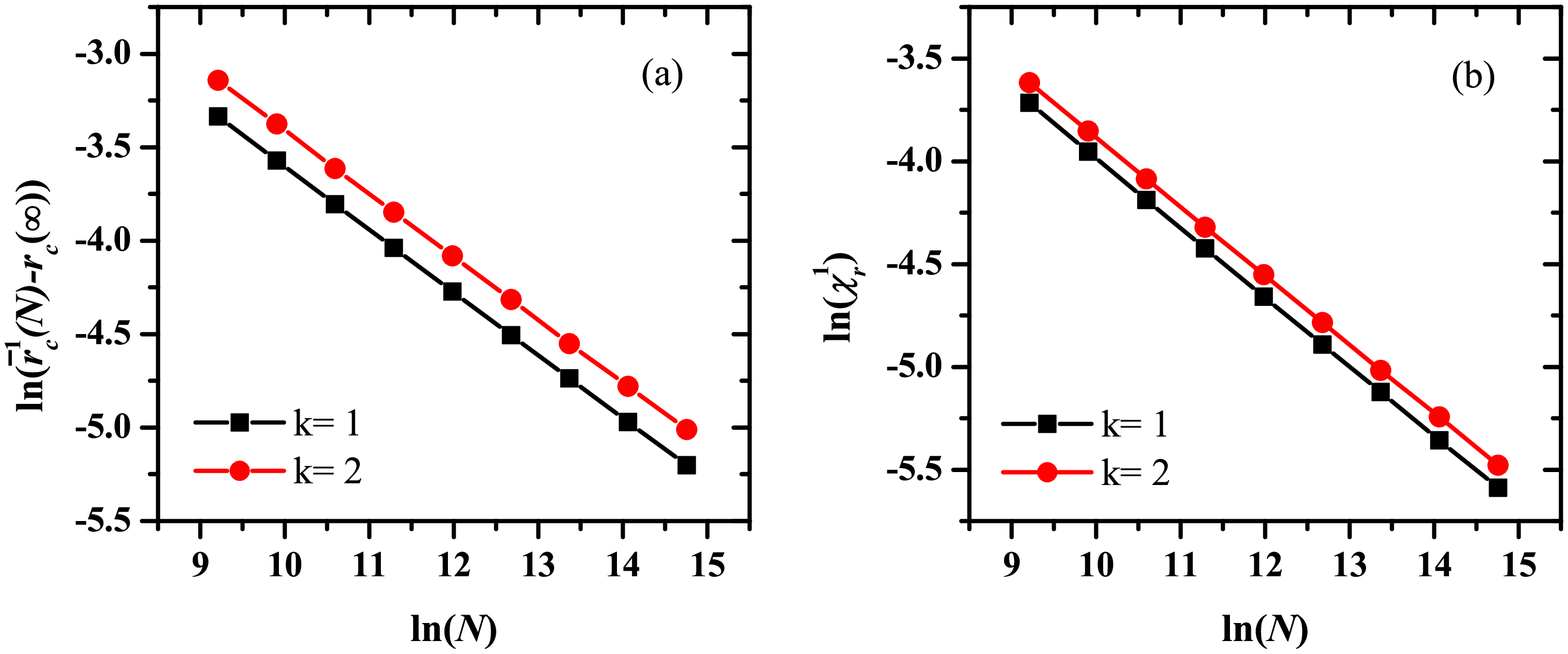}}
\caption{Asymptotic behaviors of the sample-dependent transition points associated with the largest size jump of $k$-core with $k=1,2$. (a)$\ln(\bar{r}_c^1 (N)-r_c(\infty))$ as a function of $\ln(N)$ with $r_c(\infty)=0.5$; (b) $\ln(\chi_{r}^1)$ as a function of $\ln(N)$. Both $(1/\nu)_1^1$ and $(1/\nu)_2^1$ equal approximately $1/3$ for both $k=1$ and $k=2$.}
\label{fig:3}
\end{figure}

In Fig.\ref{fig:3}, the asymptotic behaviors of the averages and fluctuations of the sample-dependent transition points associated with the largest size jump are demonstrated. $\bar{r}_c^1 (N)$ converges to the critical point $r_c(\infty)=0.5$ while $\chi_{r}^1$ converges to zero as $N\to \infty$. Linear fitting suggests that $(1/\nu)_1^1=(1/\nu)_2^1=1/3$ for $k=1,2$.

As has been demonstrated~\cite{Zhu15b}, for continuous percolation, the exponent $(1/\nu)_1^1$ obtained from fluctuations is the intrinsic exponent characterizing the divergence of correlation length. Accordingly, the exponent $\beta$ associated with order parameter can be obtained by $\beta=\frac{(\beta/\nu)_1^1}{(1/\nu)_1^1}$. For $k=1$, we get $\nu=3$ and $\beta=1$ which are the standard mean-field exponents for percolation. For $k=2$, we get $\nu=3$ and $\beta=1$ which are consistent with the results in \cite{Chalupa79}. Same results can be achieved from the second largest size jump of $k$-core.

\section{The case of hybrid $k$-core percolation ($k\ge 3$)}
\label{sec:result_k345}
When $k\ge 3$, $k$-core percolation turns into discontinuous phase transition. The results of largest and second largest size jumps of $k$-core will be presented separately since they differs a lot.

\subsection{Properties of the largest jump}
\begin{figure}[h!]
\resizebox{0.48\textwidth}{!}{\includegraphics{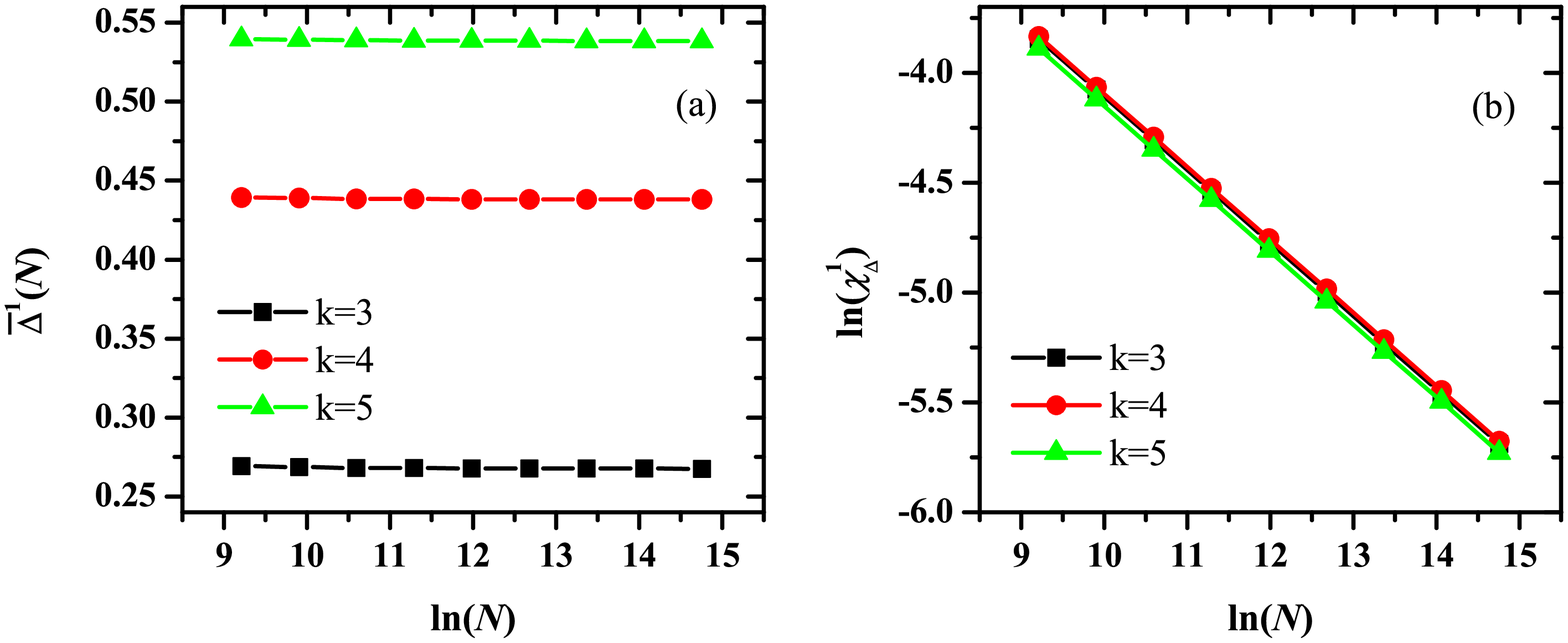}}
\caption{Asymptotic behaviors of largest size gaps of $k$-core with $k=3,4,5$. (a)$\bar{\Delta}^1 (N)$ as a function of $\ln(N)$; (b) $\ln (\chi_{\Delta}^1)$ as a function of $\ln(N)$. The exponent $(\beta/\nu)_2^1\approx 1/3$. }
\label{fig:4}
\end{figure}

As shown in Fig.\ref{fig:1}(b), when $k\ge 3$, the largest size jump of $k$-core in a single realization occurs when $k$-core emerges for the first time. In Fig.\ref{fig:4}(a), the averaged largest size gap of $k$-core with $k=3,4,5$ are plotted against $\ln(N)$. As $N$ increases, $\bar{\Delta}^1 (N)$ remains as finite constants which are about $0.2676$, $0.438$ and $0.5384$ for $k=3,4,5$ respectively. They are well in agreement with the analytical order parameter $s(r)$ as $r\to r_c$ from above reported in \cite{Pittel96}. Therefore, the size of $k$-core undergoes a finite jump in the thermodynamic limit $N\to \infty$, indicating $k$-core percolation is discontinuous. In Fig.\ref{fig:4}(b), the fluctuation of largest size gap $\chi_{\Delta}^1$ scales algebraically with system size as $N^{-1/3}$.

\begin{figure}[h!]
\resizebox{0.48\textwidth}{!}{\includegraphics{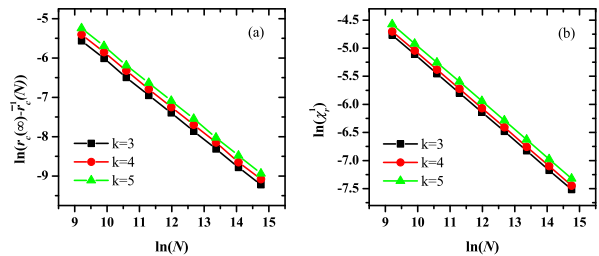}}
\caption{Asymptotic behaviors of sample-dependent transition points associated with the largest size jump of $k$-core with $k=3,4,5$. (a)$\ln(r_c(\infty)-\bar{r}_c^1 (N))$ as a function of $\ln(N)$; (b) $\ln(\chi_{r}^1)$ as a function of $\ln(N)$. The exponent $(1/\nu)_1^1$ and $(1/\nu)_2^1$ equal approximately $2/3$ and $1/2$ respectively in all the three cases.}
\label{fig:5}
\end{figure}

In Fig.\ref{fig:5}(a), the average values of sample-dependent transition points associated with the largest size jump of $k$-core are displayed. Like continuous percolation, $\bar{r}_c^1 (N)$ converges in power law to the reported analytical transition point which is $1.675459435$ for $k=3$, $2.574701375$ for $k=4$ and $3.399637745$ for $k=5$~\cite{Pittel96}. Furthermore, the obtained exponents $(1/\nu)_1^1$ for $k=3,4,5$ are all approximately equal to $2/3$ rather than $1/3$ for $k=1,2$. The corresponding fluctuations $\chi_r^1(N)$ converge to zero in power law as can be seen in Fig.\ref{fig:5}(b). The obtained exponent $(1/\nu)_2^1=0.5$ is obviously different from $(1/\nu)_1^1$.

\subsection{Properties of the second largest jump}

\begin{table*}[!ht]
\caption{Exponents of $k$-core percolation. Those with superscript 1 and 2 are obtained respectively from largest and second largest size jump behaviors of $k$-core. The last three columns are conjectured exact values.}
\begin{ruledtabular}
\begin{tabular}{c|ccccc|ccc}
$k$ & 1 & 2 & 3 & 4 & 5 & 1 & 2 & $\ge 3$ \\ \hline
$(\beta/\nu)_1^1$ & 0.331(4) & 0.662(5) & 0 & 0 & 0 & 1/3 & 2/3 & 0 \\
$(\beta/\nu)_2^1$ & 0.330(6) & 0.666(6) & 0.333(3) & 0.334(2) & 0.332(3) & 1/3 & 2/3 & 1/3 \\
$(1/\nu)_1^1$ & 0.335(2) & 0.336(3) & 0.661(6) & 0.667(9) & 0.667(9) & 1/3 & 1/3 & 2/3 \\
$(1/\nu)_2^1$ & 0.334(3) & 0.336(3) & 0.495(5) & 0.500(5) & 0.495(5) & 1/3 & 1/3 & 1/2 \\
$(\beta/\nu)_1^2$ & 0.335(2) & 0.660(6) & 0.340(8) & 0.337(5) & 0.337(4) & 1/3 & 2/3 & 1/3 \\
$(\beta/\nu)_2^2$ & 0.333(2) & 0.668(9) & 0.338(5) & 0.336(4) & 0.335(5) & 1/3 & 2/3 & 1/3 \\
$(1/\nu)_1^2$ & 0.333(2) & 0.339(5) & 0.667(4) & 0.658(8) & 0.658(6) & 1/3 & 1/3 & 2/3  \\
$(1/\nu)_2^2$ & 0.329(6) & 0.339(6) & 0.668(8) & 0.659(8) & 0.662(6) & 1/3 & 1/3 & 2/3  \\
\end{tabular}
\end{ruledtabular}
\label{table1}
\end{table*}

\begin{figure}[h!]
\resizebox{0.48\textwidth}{!}{\includegraphics{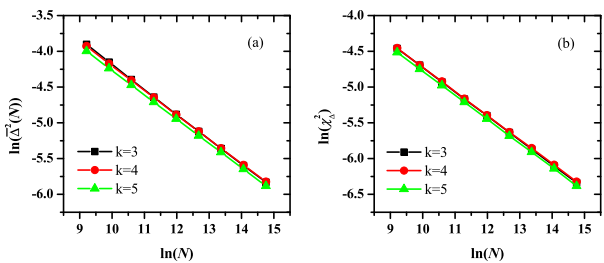}}
\caption{Asymptotic behaviors of the second largest size jump of $k$-core with $k=3,4,5$. (a)$\ln(\bar{\Delta}^2 (N))$ as a function of $\ln(N)$; (b) $\ln(\chi_{\Delta}^2)$ as a function of $\ln(N)$. Both the exponent $(\beta/\nu)_1^2$ and $(\beta/\nu)_2^2$ equal approximately $1/3$. }
\label{fig:6}
\end{figure}

The second largest size jump in each run is much smaller than the largest one as shown in Fig.\ref{fig:1}(b).  The asymptotic behaviors of average values and fluctuations of second largest size jump of $k$-core with $k=3,4,5$ are illustrated in Fig.\ref{fig:6}. In contrast to the largest size jump, $\bar{\Delta}^2 (N)$ shows power-law convergence to zero which is the typical characteristic of continuous percolation. The corresponding fluctuations $\chi_{\Delta}^2$ also converges to zero in power law. Both the exponents $(\beta/\nu)_1^2$ and $(\beta/\nu)_2^2$ are approximately $1/3$.

\begin{figure}[h!]
\resizebox{0.48\textwidth}{!}{\includegraphics{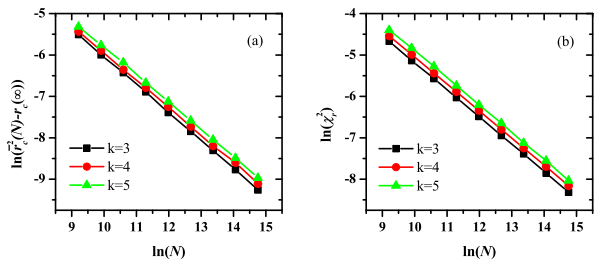}}
\caption{Asymptotic behaviors of sample-dependent transition points associated with the second largest size jump of $k$-core with $k=3,4,5$. (a)$\ln(\bar{r}_c^2 (N)-r_c(\infty))$ as a function of $\ln(N)$; (b) $\ln(\chi_r^2(N))$ as a function of $\ln(N)$. Both the exponent $(1/\nu)_1^2$ and$(1/\nu)_2^2$ equal approximately $2/3$.}
\label{fig:7}
\end{figure}

In Fig.\ref{fig:7}(a), we plot the $\ln(\bar{r}_c^2 (N)-r_c(\infty))$ against $\ln(N)$ with the same $r_c(\infty)$ as in Fig.\ref{fig:5}(a). Three parallel straight lines are displayed in the ln-ln scale. Thus, $\bar{r}_c^2 (N)$ converges in power-law to the the same transition point as $\bar{r}_c^1 (N)$ does but in the opposite direction. However, the exponent $(1/\nu)_1^2$ is about $2/3$ other $1/3$.

Since the second largest size jump of $k$-core occurs after the largest one in the process of adding edges to networks, the fluctuations of transition points from second largest size jump are connected with that of the largest one. To eliminate this effect, we redefine $\chi_r^2$ as the fluctuation of shift of $r$ between the largest and second largest size jumps occurs in each run. Explicitly,
\begin{equation}
\chi_r^2\equiv \sqrt{<(r_c^{2(i)}-r_c^{1(i)})^2>-<r_c^{2(i)}-r_c^{1(i)}>^2}
\end{equation}

Unlike the case of largest size jump where $\chi_r^1$ converges with a different exponent from $\bar{r}_c^1$, $\chi_r^2$ displays power law decay with exponent $(1/\nu)_2^2\approx 2/3$ which agrees with $(1/\nu)_1^2$ of $\bar{r}_c^2$.

All the exponents of $k$-core percolation in ER random networks are summarized in Table.\ref{table1}. When $k\ge 3$, all the exponents are nearly independent of $k$ and their conjectured exact values are given in the last column. Particularly, from $(\beta/\nu)_1^2$ and $(1/\nu)_1^2$, we can get
\begin{equation}
\beta=\frac{(\beta/\nu)_1^2}{(1/\nu)_1^2}=0.5
\end{equation}
which is coincident with the exponent in Eq.\eqref{eq:hybrid}. In addition, same results can be obtained from third, forth...largest size jump behaviors of $k$-core.

\section{Summary}
\label{sec:summary}
We have studied the $k$-core percolation in ER networks through analysis of size jump behaviors of $k$-core during the process of adding edge to networks. It's confirmed that $k$-core percolation is continuous when $k=1,2$ and discontinuous but with the characteristic of continuous phase transition when $k\ge 3$. In other words, $k$-core percolation is a hybrid first-order-second-order phase transition when $k\ge 3$.

When $k=1,2$, the largest and second largest size jumps possess the same asymptotic properties as $N\to \infty$. The $1$-core percolation, namely percolation of giant component, belongs to the universality class of  mean-field values of percolation in infinite dimension where $\nu=3$ and $\beta=1$.  The $2$-core percolation belongs to a different universality class where $\nu=3$ and $\beta=2$. There are some other percolation models sharing the same universality class, such as percolation of bi-connected component in undirected networks, percolation of giant strongly connected component, giant in- and out-component in directed networks.

When $k\ge 3$, the asymptotic behaviors of the largest size jump reflect the discontinuity of $k$-core percolation while the other smaller size jumps contain the properties of continuous percolation. Despite its hybrid nature, we obtain a set of exponents which are independent of $k$.

\end{document}